\documentclass[aps,prl,superscriptaddress,showpacs,floatfix]{revtex4-1}
\usepackage{graphicx,amsmath,amssymb,xspace,epsfig,float,multirow,subfigure,tabularx}
\usepackage{amsbsy,amssymb,amsmath,bm}
\usepackage{epsf}
\usepackage{color,natbib}
\usepackage{amsfonts}
\usepackage{hyperref}

\pdfoutput=1

\begin{document}

\title{Two step I to II type transitions in layered Weyl semi-metals and
their impact on superconductivity}
\author{Baruch Rosenstein}
\email{vortexbar@yahoo.com}
\author{B.Ya.Shapiro}
\email{shapib@mail.biu.ac.il}

\affiliation{Department of Electrohysics, National Yang Ming Chiao Tung University, Hsinchu,
Taiwan, R.O.C. }

\affiliation{Department of Physics, Institute of Superconductivity, Bar-Ilan
University, 52900 Ramat-Gan, Israel.}

\begin{abstract}
Novel "quasi two dimensional" typically layered (semi) metals offer a unique
opportunity to control the density and even the topology of the electronic
matter. Along with doping and gate voltage, a robust tuning is achieved by
application of the hydrostatic pressure. In Weyl semi - metals the tilt of
the dispersion relation cones, $\kappa ,$ increases with pressure, so that
one is able to reach type II ($\kappa >1$starting from the more conventional
type I Weyl semi - metals.$\kappa <1$. The microscopic theory of such a
transition is constructed. It is found that upon increasing pressure the I
to II transition occurs in two continuous steps. In the first step the cones
of opposite chirality coalesce so that the chiral symmetry is restored,
while the second transition to the Fermi surface extending throughout the
Brillouin zone occurs at higher pressures. Flattening \ of the band leads to
profound changes in Coulomb screening. Superconductivity observed recently
in wide range of pressure and chemical composition in Weyl semi-metals of
both types. The phonon theory of pairing including the Coulomb repulsion for
a layered material is constructed and applied to recent extensive
experiments on $HfTe_{5}$.
\end{abstract}

\pacs{74.20.Fg, 74.70.-b, 74.62.Fj}
\maketitle

\section{Introduction.}

\textit{\ }The 3D and 2D topological quantum materials, such as topological
insulators and Weyl semi - metals (WSM), attracted much interests due to
their rich physics and promising prospects for applications. The band
structure in the so called type I WSM like graphene\cite{Katzenelson} in 2D
and $ZrTe_{5}$ in 3D \cite{Weng,MoTeearly,ZrTe}\cite{Armitage}, is
characterized by appearance of linear dispersion relation, cones around
several Dirac points, due to the band inversion. This is qualitatively
distinct from conventional metals, semi - metals or semiconductors, in which
bands are typically parabolic. The dispersion cones are often tilted\cite%
{Goerbig}. In an extreme case of type-II WSMs, the cones have such a strong
tilt, $\kappa \geq 1$, that they exhibit a nearly flat band at Fermi surface
first predicted\cite{Soluyanov} in $WTe_{2}$. Typically the Fermi surface
"encircles" the Brillouin zone and therefore is topologically distinct from
conventional "pockets". This in turn leads to exotic electronic properties
different from conventional and the type I materials. Examples include the
collapse of the Landau level spectrum in magnetoresistance \cite{Yu}, and
novel quantum oscillations \cite{Brien}. Several \textit{layered }materials
were predicted and observed to undergo\cite{toptransition} the I to II
(abbreviated as $I\rightarrow II$) transition while doping or pressure is
changed \cite{CdAs}\cite{NaBi}. In fact a well known layered organic
compound $\alpha -(BEDT-TTF)_{2}I_{3}$ was a long time suspected \cite%
{Goerbig} to be a quasi - 2D materials undergoing such transition.

Recent experiments concentrated on two (close) families of layered
materials. The first is superlattice of transition metal dichalcogenides 
\cite{Wang}layers with formula $MX_{2}$. The metals include $M=Mo,W,V,Ta,Pd$%
, and the chalcogenides $X=S,Te,Se$. Majority of representatives of these
class are 2D WSM. The well separated layers are integrated into van der
Waals heterostructures by vertically stacking \cite{Britnel}\cite{TaX2}.
Intercalation and external pressure are the direct and effective methods for
achieving exotic properties distinctive from the pristine materials\cite%
{pressureTc}\cite{PdTe2}. Yet another class of stacked transition metal
pentatellurides, including $HfTe_{5}$ and $ZrTe_{5}$, were recently
comprehensively investigated \cite{ZrTe} \cite{Liu17}.\ For example the
transport and superconductive properties of $HfTe_{5}$ were comprehensively
studied \cite{Liu17} at pressures as high as $30GPa$.

Pressure in particular\cite{Kusmartsev} controls both the strength of the
interlayer coupling and of the cone slope allows to observe the topological
transition. The affect on physical properties of the topological phase
transitions between the type I to type II Weyl phases was considered
theoretically. In ref.\cite{Yeh} the heat capacity, compressibility and
magnetic susceptibility was studied. Superconductivity observed recently in
wide range of pressure and chemical composition in Weyl semi-metals of both
types. In \ the previous paper\cite{Rosenstein17} and a related work\cite%
{Zyuzin} a continuum theory of conventional superconductivity through the $%
I\rightarrow II$ topological transition was developed. Magnetic response in
the superconducting state were calculated in \cite{Rosenstein18}\cite{GL18}.
The continuum approach used was too "mesoscopic" in order to describe the
transition region since the global topology of the Brillouin zone is beyond
the scope of the continuum approach.

In the present paper a theory of the topological transitions of the electron
liquid of layered WSM under hydrostatic pressure is constructed using a
(microscopic) tight binding model on the honeycomb lattice similar to that
used to model \cite{Li15} dichalcogetite $2H$ $WTe_{2}$. It possesses an
important chiral symmetry between two Brave (hexagonal) sublattices. The
Weyl cones of opposite chirality appear at the crystallographic $K$ and $%
K^{\prime }$ points for $\kappa =0$. The (discrete) chiral symmetry persists
at all values of $\kappa $. This relatively simple model describes well both
classes of layered materials that are Weyl semimetals.

Unexpectedly investigation of the pressure - "topology" phase diagram of
this sufficiently universal microscopic model reveals that (at nonzero
chemical potential) the $I\rightarrow II$ transition always occurs in two
steps. In the first step upon increasing pressure leading to higher tilt $%
\kappa $ the circular pockets around the cones of opposite chirality
coalesce into a single (type I) elliptic Fermi surface. The chiral symmetry
chiral symmetry is spontaneously broken. The second transition to the type
II Fermi surface (extending throughout the Brillouin zone) occurs at yet
higher pressures.

As in previous investigations\cite{Rosenstein17}\cite{Zyuzin}
superconductivity is used as an efficient signature of the topological
transition. The phonon pairing theory was improved compared to previous work
by accounting for the effects of screened Coulomb repulsion. We calculate
the superconducting critical temperature taking into consideration the
modification of the Coulomb electron-electron interaction. The Gorkov
equations for two sublattics system are solved without resorting to the
mesoscopic approach. Moreover it turns out that the screening of Coulomb
repulsion plays a much more profound role in quasi 2D materials and do not
allow the pseudo-potential simplification developed by MacMillan\cite%
{McMillan}. Taking this into account involves a nontrivial dependence on
quasi-momentum in the gap equation (along with frequency dependence). The
results compare well with recent experiment on\cite{Liu17} $HfTe_{5}$.

Rest of the paper is organized as follows. In Section II the universal
microscopic model of the layered WSM is described. The dependence of the
tilt parameter $\kappa $, electron density and the interlayer distance on
pressure are phenomenologically related to parameters of the model. In
Section III the Gorkov equations for the optical phonon mediated intra -
layer pairing for a multiband system including the Coulomb repulsion is
derived and solved numerically. In Section IV the phonon theory of pairing
including the Coulomb repulsion for a layered material is applied to recent
extensive experiments on $HfTe_{5}$ under the hydrostatic pressure. The last
Section contains conclusions and discussion.

\section{A "universal" lattice model of layered (type I and type II) Weyl
semi-metals}

\subsection{Inter - layer hopping on honeycomb lattice}

A great variety of tight binding models were used to describe Weyl (Dirac)
semimetals in 2D. Historically the first was graphene (type I, $\kappa =0$)
, in which electrons hope between the neighboring cites of the honeycomb
lattice. Two Dirac cones appear at $K$ and $K^{\prime }$ crystallographic
points in Brillouin one (BZ). Upon modification (gate voltage,
pressure,intercalation) the hexagonal symmetry is lost, however a discrete
chiral symmetry between two sublattices, denoted by $I=A,B$, ensures the 2D
WSM. The tilted type I and even type II ($\kappa >1$) WSM can be described
by the same Hamiltonian with the tilt term added. We restrict the discussion
to systems with the minimal two cones of opposite chirality and negligible
spin orbit coupling. This model describes the compounds listed in
Introduction and can be generalizable to more complicated WSM. This 2D model
is extended to a layered system with interlayer distance $d$. The 2D WSM
layers are separated by dielectric streaks with interlayer hopping
neglected, so that they are coupled electromagnetically only\cite{Elliasson}.

The lateral atomic coordinates on the honeycomb lattice are $\mathbf{r}_{%
\mathbf{n}}=n_{1}\mathbf{a}_{1}+n_{2}\mathbf{a}_{2}$, where lattice vectors
are: 
\begin{equation}
\mathbf{a}_{1}=a\left( \frac{1}{2},\frac{\sqrt{3}}{2}\right) ;\text{ }%
\mathbf{a}_{2}=a\left( \frac{1}{2},-\frac{\sqrt{3}}{2}\right) \text{.}
\label{unit cell}
\end{equation}%
The length of the lattice vectors $a$ will be taken as the length unit and
we also set $\hbar =1$. The hopping Hamiltonian including the tilt term is:

\begin{equation}
K=\sum \nolimits_{\mathbf{n}l}\left \{ t\left( \sum \limits_{i=1,2,3}\psi _{%
\mathbf{n}l}^{sA\dagger }\psi _{\mathbf{r}_{\mathbf{n}}+\mathbf{\delta }%
_{i},l}^{sB}+\mathrm{h.c.}\right) -\kappa \psi _{\mathbf{n}l}^{sI\dagger
}\psi _{\mathbf{r}_{\mathbf{n}}+\mathbf{a}_{1},l}^{sI}-\mu n_{\mathbf{n}%
,l}\right \} \text{.}  \label{Energy}
\end{equation}%
Here\ an integer $l$ labels the layers. Operator $\psi _{\mathbf{n}%
l}^{sA\dagger }$ is the creation operators with spin $s=\uparrow ,\downarrow 
$, while the density operator is defined as $n_{\mathbf{n}l}=\psi _{\mathbf{n%
}l}^{sI\dagger }\psi _{\mathbf{n}l}^{sI}$. The chemical potential is $\mu $,
while $t$ is the hopping energy. Each site has three neighbors separated by
vectors $\mathbf{\delta }_{1}=\frac{1}{3}\left( \mathbf{a}_{1}-\mathbf{a}%
_{2}\right) ,\mathbf{\delta }_{2}=-\frac{1}{3}\left( 2\mathbf{a}_{1}+\mathbf{%
a}_{2}\right) $ and $\mathbf{\delta }_{3}=\frac{1}{3}\left( \mathbf{a}_{1}+2%
\mathbf{a}_{2}\right) $. Dimensionless parameter $\kappa $ determines the
tilt of the Dirac cones along the $\mathbf{a}_{1}$direction\cite{Goerbig}.
In the 2D Fourier space, $\psi _{n_{1}n_{2}l}^{sA\dagger }=N_{s}^{-2}\sum
\nolimits_{k_{1}k_{2}}\psi _{k_{1}k_{2}l}^{sA\dagger }\exp \left[ 2\pi
i\left( k_{1}n_{1}\mathbf{+}k_{2}n_{2}\right) /N_{s}\right] $, one obtains
for Hamiltonian \bigskip (for finite discrete reciprocal lattice $%
N_{s}\times N_{s}$):

\begin{equation}
K=\frac{1}{N_{s}^{2}}\sum\nolimits_{k_{1}k_{2}l}\psi
_{k_{1}k_{2}l}^{s\dagger }M_{k_{1}k_{2}}\psi _{k_{1}k_{2}l}^{s}\text{.}
\label{Fourier space}
\end{equation}%
Here $\mathbf{k}=\frac{k_{1}}{N_{s}}\mathbf{b}_{1}+\frac{k_{2}}{N_{s}}%
\mathbf{b}_{2}$ are the reciprocal lattice vectors and the matrix 
\begin{equation}
M_{\mathbf{k}}=d_{\mathbf{k}}^{x}\sigma _{x}+d_{\mathbf{k}}^{y}\sigma
_{y}+d_{\mathbf{k}}^{0}I  \label{Matrix}
\end{equation}
where 
\begin{eqnarray}
d_{\mathbf{k}}^{x} &=&\cos \left[ \frac{2\pi }{3N_{s}}\left(
k_{1}-k_{2}\right) \right] +2\cos \left[ \frac{\pi }{N_{s}}\left(
k_{1}+k_{2}\right) \right] \cos \left[ -\frac{\pi }{3N_{s}}\left(
k_{1}-k_{2}\right) \right] ;  \label{M} \\
d_{\mathbf{k}}^{y} &=&-\sin \left[ \frac{2\pi }{3N_{s}}\left(
k_{1}-k_{2}\right) \right] +2\cos \left[ \frac{\pi }{N_{s}}\left(
k_{1}+k_{2}\right) \right] \sin \left[ \frac{\pi }{3N_{s}}\left(
k_{1}-k_{2}\right) \right] ;  \notag \\
d_{\mathbf{k}}^{0} &=&-\kappa \cos \left[ \frac{2\pi }{N_{s}}k_{1}\right]
-\mu \text{.}  \notag
\end{eqnarray}%
From now on the hopping energy $t$ will be our energy unit.

The free electrons part of the Matsubara action for Grassmanian fields $\psi
_{\mathbf{k}ln}^{\ast sI}$ therefore is:

\begin{equation}
S^{e}=\frac{1}{T}\sum \nolimits_{\mathbf{k}ln}\psi _{\mathbf{k}ln}^{\ast
sA}\left \{ \left( -i\omega _{n}+d_{\mathbf{k}}^{0}\right) \delta
^{AB}+\sigma _{i}^{AB}d_{\mathbf{k}}^{i}\right \} \psi _{\mathbf{k}ln}^{sB}%
\text{.}  \label{Action_e}
\end{equation}%
Here $\omega _{n}=\pi T\left( 2n+1\right) $ is the Matsubara frequency. The
Greens' function, $g_{\mathbf{k}n}^{ss^{\prime }}=\delta ^{ss^{\prime }}g_{%
\mathbf{k}n}$, of free electrons has the (sublattice) following matrix form:

\begin{equation}
g_{\mathbf{k}n}=\left[ \left( -i\omega _{n}+d_{\mathbf{k}}^{0}\right)
I+\sigma _{i}d_{\mathbf{k}}^{i}\right] ^{-1}=\frac{\left( -i\omega _{n}+d_{%
\mathbf{k}}^{0}\right) I-\sigma _{i}d_{\mathbf{k}}^{i}}{\left( i\omega
_{n}-d_{\mathbf{k}}^{0}\right) ^{2}-\left( d_{\mathbf{k}}^{x2}+d_{\mathbf{k}%
}^{y2}\right) }\text{.}  \label{gdef}
\end{equation}%
Now we turn to the interactions part of the Hamiltonian.

\subsection{Coulomb repulsion}

The electron-electron repulsion in the layered WSM on the lattice can be
presented in the form,

\begin{equation}
V=\frac{e^{2}}{2}\sum \nolimits_{\mathbf{nn}^{\prime }ll^{\prime }}n_{%
\mathbf{n}l}v_{\mathbf{n-n}^{\prime },l-l^{\prime }}^{C}n_{\mathbf{n}%
^{\prime }l^{\prime }}\text{,}  \label{Coulomb}
\end{equation}%
where $v_{\mathbf{n-n}^{\prime },l-l^{\prime }}^{C}$ is the "bare" Coulomb
interaction between electrons. Making the 2D Fourier transform, one obtains,%
\begin{equation}
V=\frac{e^{2}}{2N_{s}^{2}}\sum \nolimits_{\mathbf{q}ll^{\prime }}n_{\mathbf{q%
}l}v_{\mathbf{q,}l-l^{\prime }}^{C}n_{-\mathbf{q}l^{\prime }}\text{,}
\label{V}
\end{equation}%
where 
\begin{equation}
v_{\mathbf{q},l-l^{\prime }}^{C}=v_{\mathbf{q}}^{2D}e^{-dq\left \vert
l-l^{\prime }\right \vert }\text{,}  \label{vC}
\end{equation}%
with the in plane Coulomb repulsion being $v_{\mathbf{q}}^{2D}=\frac{2\pi
e^{2}}{q\epsilon }$. Here $\epsilon $ is the inter - layer dielectric
constant \cite{dielectric}, while $d$ is the interlayer distance. \bigskip
On the hexagonal lattice the exponential formula approximates the Coulomb
repulsion well only away from the BZ boundaries. Near the boundaries the
(periodic) potential is calculated numerically in SI3. The long range
screening effect of the Coulomb interaction is effectively taken into
account using the RPA approximation. Effect of pressure on the various
parameters is discussed in the next section.

\section{Two step I to II type topological transition}

\subsection{Pressure induced parameter modifications}

While pressure turned out to be more experimentally accessible control
parameter than the gate voltage, in the early works mentioned in
Introduction typically the phase diagram was studied as a function of the
chemical potential. Moreover in most recent experiments the hydrostatic
pressure serves as a control parameter to induce topological transformations
of the electronic matter in WSM. The parameter dependence of a microscopic
model on pressure, is in principle derivable by the DFT and a corresponding
adaptation of the elasticity theory \cite{Kusmartsev}. Although there exist
a qualitative theoretical description of the pressure dependence of the
Coulomb repulsion\cite{Goerbig2}, electron-phonon coupling and the topology
of the Fermi surface of these novel materials \cite{Sun}, it is difficult to
determine quantitatively the tilt $\kappa $, inter layer spacing $d$,
electron density and other parameters. Therefore we use an experimentally
parametrized (see for example a comprehensive study\cite{MoTe2}) dependence
of these parameters on the pressure. In the present paper to describe a
specific material $HfTe_{5}$ as an example we utilize experimental results
of ref.\cite{Liu17}. Note that in many materials the robust electron gas
exists only at certain pressure.

For not very large pressures ($P<15$ $GPa$) several parameters dependencies
can be accounted for as linear. In particular, the layer spacing are the
tilt parameter are modified under pressure $P$ as: 
\begin{eqnarray}
d\left( P\right) &=&\frac{d_{a}}{1+\sigma P/d_{a}}\approx d_{a}-\sigma \text{
}P\text{;}  \label{d(P)} \\
\kappa \left( P\right) &=&\kappa _{a}+\gamma P\text{.}  \notag
\end{eqnarray}
The tilt parameter was estimated in ref.\cite{Kusmartsev} for a wide range
of $\kappa $. For layered $HfTe_{5}$ the stress parameter is $\sigma
=0.225A/GPa$. The "ambient" value is $d_{a}=7.7A$. As noted above the
electron gas exists\cite{Liu17} in this case only for $P>3GPa$. For the tilt
modulus $\kappa _{a}=-0.3$ and $\gamma =0.15/GPa$.

Measurements demonstrate that 3D electron density in the type I phase of
layered WSM is exponential in pressure (for not very high pressures): 
\begin{equation}
n^{3D}\left( P\right) =n_{a}e^{\beta P}\text{.}  \notag
\end{equation}%
It saturates upon approach to type II WSM. The ambient value is $%
n_{a}=1.4\times 10^{19}cm^{-3}$, while $\beta =0.77/GPa$. The two
dimensional electron density in the layers is related to the measured
density by $n\left( P\right) =n^{3D}\left( P\right) d\left( P\right) $. The
influence on the interactions will be discussed in the next Section. Having
described the model let us turn to the spectrum and topology of the Fermi
surface for different pressures.

\begin{figure}[h]
\centering \includegraphics[width=18cm]{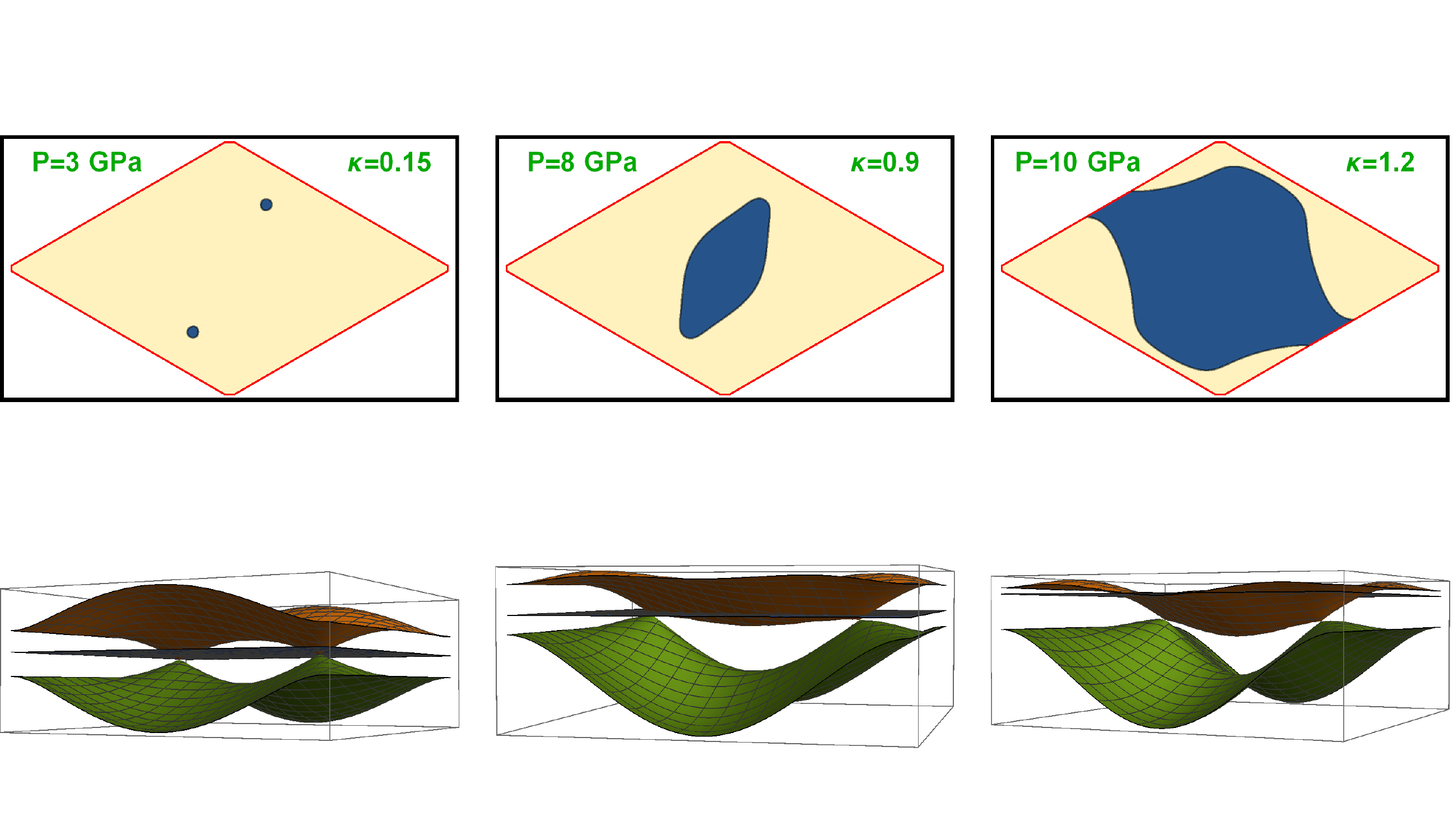}
\caption{Evolution of the Fermi surface topology as the pressure of Weyl
semimetal increase. Parameters like the tilt $\protect\kappa \left( P\right) 
$, electron density etc are given in Eqs.(\protect\ref{d(P)}) of the Weyl
semimetal. \ The upper raw depicts the Fermi surfaces of all three
topological phases, while the lower row are the corresponding dispersion
relation of both branches (brown and green surfaces) with respect to Fermi
level (the blue plane). At relatively low pressure the FS consists of two
small Dirac pockets. At intermediate pressures the two pockets merge into a
single ellipsoidal large pocket (still type I). At very high pressures the
electron liquid undergoes the type I to type II topological transition.}
\end{figure}

\bigskip

\subsection{Topological phases of layered WSM}

Upon increasing pressure the I to II transition occurs in two continuous
steps. In the first step the cones of opposite chirality coalesce so that
the chiral symmetry is restored, while the second transition to the Fermi
surface extending throughout the Brillouin zone occurs at higher pressures.
Fig.1 describes the Fermi surface (blue areas depict the Fermi sea in upper
contour plots) and dispersion relations (lower 3D plots) of three
representative pressures value from the three phases. There are two branches
(brown higher than green) crossing the Fermi level (blue plane).

The graphene - like dispersion relation for smallest value of pressure when
the electron pockets exist, $P=3$ $GPa$, $\kappa =0.15$ (left panel in
Fig.1) represents the type I WSM below the chiral transition. A rhombic BZ
(with coordinates $k_{1}$ and $k_{2}$ defined in Eq.(\ref{Fourier space}),
yellow area covers the BZ) is chosen. Location of the cones (see a lower 3D
plot) are close to crystallographic $K^{\pm }$ points. There are two
slightly tilted Dirac cones of opposite chirality. Increasing the pressure
towards the chiral transition (see more plots in SI2) at $P_{\chi }=6.5$ $%
GPa $, the two pockets of the Fermi surface become elongated and larger and
eventually merge into a single pocket shown in the central figure for $P=8$ $%
GPa$. The tilt parameter is already significant $\kappa =0.9$. At yet larger
pressure $P=8$ $GPa$ (right panel) the material becomes a type II WSM with
large $\kappa >1.2$. In this case FS envelops the BZ that topologically is
torus. See the segment on the boundary $k_{2}=0=2\pi /a$. Obviously the
upper band becomes flatter as the tilt (pressure) increases.

Fig.2 gives the 2D electron density and the density of states as function of
the chemical potential for Hamiltonian of the previous Section.

\begin{figure}[h]
\centering \includegraphics[width=14cm]{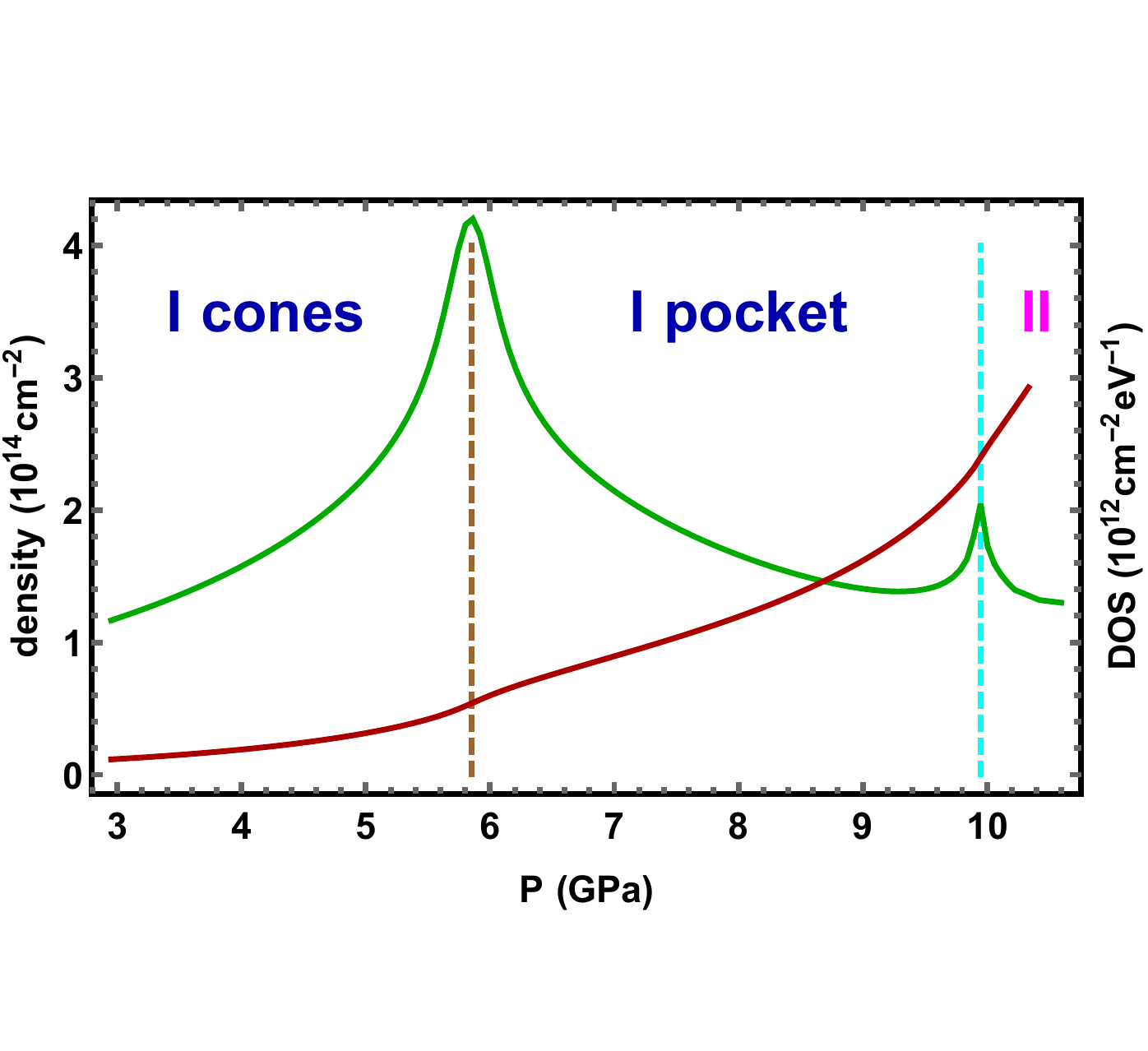}
\caption{Electron density and density of states (DOS) as function of
pressure $P$.of WSM. The 2D electron density (the brown curve) monotonically
increases, while DOS (the green curve) has cusps at both topological
transitions. On the cusp the derivative of DOS with respect pressure changes
sign.}
\end{figure}
Both the electron density and the density of states were calculated
numerically for the Fermi distribution function at temperature $T=1K$ (the
density at zero temperature corresponds to an area inside the FS) at various
values of the chemical potential. Then the density is matched with those
determined phenomenologically in the previous subsection.

\subsection{The first topological transition: spontaneous chiral symmetry
breaking}

At small pressures, $3$ $GPa<P<4$ $GPa$, the Fermi surface consists of two
well separated Dirac cones of opposite chirality. The tilt does not affect
the basic chiral symmetry of the honeycomb lattice: two sublattices are
related by a reflection. The sixfold symmetry in undistorted graphene is of
course typically broken down to the reflection symmetry only. When the
tilted cones FS pockets merge at the transition \ $P=P_{\chi }=6.5$ $GPa$
(see the brown line in Fig.2) the chiral symmetry of the ground state is
restored. The overall chirality of the FS above $P_{\chi }$ (a topological
number) therefore is zero. Although we are not aware of a mathematical
proof, this transition always precedes the $I\rightarrow II$ topological
transition, see cyan line in Fig.2. The chiral transition is also
topological, but a more local sense: fracture of the Fermi surface like in
graphene oxide\cite{grapheneoxide} or Lifshitz transition in high $T_{c}$
cuprates like $La_{2-x}Sr_{x}CuO_{4}$. The $I\rightarrow II$ is more "exotic"%
\cite{Volovik}: it involves the global topology of the Fermi surface (it is
a torus). The DOS at transition \ (the green curve in Fig.2) has a finite
maximum at which the derivative changes sign.

\subsection{The second topological transition: $I\rightarrow II$}

The electron density in type I phase above the chiral transition grows quite
fast, see red line in Fig.2, so that at large pressures a significant part
of BZ for one of the branches of spectrum is occupied. Eventually at $%
P_{I\rightarrow II}=9.9$ $GPa$ the growing single pocket envelops the BZ
torus and thus FS splits again into two curves, see the right panel in
Fig.1. Density of electron saturates, while the DOS has another finite peak.
The two transition lead to singularities in various physical quantities. In
the next Section the screening of Coulomb interactions is discussed.

\section{Screening in layered Weyl semi - metal.}

The screening in the layered system can be conveniently partitioned into the
screening within each layer described by the polarization function $\Pi _{%
\mathbf{q}n}$ and electrostatic coupling to carriers in other layers. We
start with the former.

\subsection{Polarization function of the electron gas in Layered WSM}

In a simple Fermi theory of the electron gas in normal state with Coulomb
interaction between the electrons in RPA approximation the Matsubara
polarization is calculated as a simple \textit{minus} "fish" diagram \cite%
{Elliasson} in the form:

\begin{equation}
\Pi _{\mathbf{q}n}=2T\sum\nolimits_{\mathbf{p}m}\text{Tr}\left[ g_{\mathbf{p}%
m}g_{\mathbf{p+q},m+n}^{tr}\right] \text{.}  \label{Polarization}
\end{equation}%
Using the GF (see Eq.(\ref{gdef})), one obtain:

\begin{equation}
\Pi _{\mathbf{q}n}=\frac{4T}{N_{s}^{2}}\sum \nolimits_{\mathbf{p}m}\frac{%
\left( i\omega _{m}+A\right) \left( i\omega _{m}+B\right) +C}{\left[ \left(
i\omega _{m}+A\right) ^{2}-\alpha ^{2}\right] \left[ \left( i\omega
_{m}+B\right) ^{2}-\beta ^{2}\right] }\text{,}  \label{Polarization omega}
\end{equation}%
where

\begin{eqnarray}
A &=&-d_{\mathbf{p}}^{0};B=i\omega _{n}-d_{\mathbf{p+q}}^{0};C=d_{\mathbf{p}%
}^{x}d_{\mathbf{p+q}}^{x}-d_{\mathbf{p}}^{y}d_{\mathbf{p+q}}^{y}
\label{Adef} \\
\alpha ^{2} &=&d_{\mathbf{p}}^{x2}+d_{\mathbf{p}}^{y2};\beta ^{2}=d_{\mathbf{%
p+q}}^{x2}+d_{\mathbf{p+q}}^{y2}\text{.}  \notag
\end{eqnarray}%
Performing summation over $m$, one obtains:

\begin{equation}
\Pi _{\mathbf{q}n}=-\frac{1}{N_{s}^{2}}\sum \nolimits_{\mathbf{p}}\left \{ 
\begin{array}{c}
\frac{\alpha ^{2}-\alpha (A-B)+C}{\alpha \left[ \left( A-B-\alpha \right)
^{2}-\beta ^{2}\right] }\tanh \frac{\alpha -A}{2T}+\frac{a^{2}+\alpha (A-B)+C%
}{\alpha \left[ \left( A-B+\alpha \right) ^{2}-\beta ^{2}\right] }\tanh 
\frac{\alpha +A}{2T} \\ 
+\frac{\beta ^{2}+\beta \left( A-B\right) +C}{\beta \left[ \left( A-B+\beta
\right) ^{2}-\alpha ^{2}\right] }\tanh \frac{\beta -B}{2T}+\frac{\beta
^{2}-\beta \left( A-B\right) +C}{\beta \left[ \left( A-B-\beta \right)
^{2}-\alpha ^{2}\right] }\tanh \frac{\beta +B}{2T}%
\end{array}%
\right \} \text{.}  \label{polarization}
\end{equation}%
The polarization function however is strongly differ from the usual Lindhard
expression for a parabolic band.

\subsection{Screening due to electron gas in layered system}

Coulomb repulsion between electrons in different layers $l$ and $l^{\prime }$
within the RPA approximation is determined by the following integral
equation:

\begin{equation}
V_{\mathbf{q,}l-l^{\prime }\mathbf{,}n}^{RPA}=v_{\mathbf{q},l-l^{\prime
}}^{C}+\Pi _{\mathbf{q}n}\sum \nolimits_{l^{\prime \prime }}v_{\mathbf{q}%
,l-l^{\prime \prime }}^{C}V_{\mathbf{q},l^{\prime \prime }-l^{\prime }%
\mathbf{,}n}^{RPA}.  \label{Series}
\end{equation}%
The polarization function $\Pi _{\mathbf{q}n}$ in 2D was calculated in the
previous subsection. This set of equations is decoupled by the Fourier
transform in the $z$ direction:

\begin{equation}
V_{\mathbf{q,}q_{z},n}^{RPA}=\frac{v_{\mathbf{q},q_{z}}^{C}}{1-\Pi _{\mathbf{%
q}n}v_{\mathbf{q},q_{z}}^{C}}\text{ , \ }  \label{RPA}
\end{equation}%
where%
\begin{equation}
v_{\mathbf{q},q_{z}}^{C}=\sum\nolimits_{l}v_{\mathbf{q}}^{2D}e^{iq_{z}l-qd%
\left\vert l\right\vert }=v_{\mathbf{q}}^{2D}\frac{\sinh \left( qd\right) }{%
\cosh \left( qd\right) -\cos \left( dq_{z}\right) }\text{.}  \label{vCdef}
\end{equation}%
The screened interaction in a single layer therefore is is given by the
inverse Fourier transform \cite{Elliasson}:

\begin{equation}
V_{\mathbf{q,}l-l^{\prime },n}^{RPA}=\frac{d}{2\pi }\int_{q_{z}=-\pi
/d}^{\pi /d}e^{iq_{z}d\left( l-l^{\prime }\right) }\frac{v_{\mathbf{q}%
q_{z}}^{C}}{1-\Pi _{\mathbf{q}n}v_{\mathbf{q}q_{z}}^{C}}\text{.}
\label{Screening}
\end{equation}%
Considering screened Coulomb potential at the same layer $l=l^{\prime },\ $%
the integration gives,

\begin{equation}
V_{\mathbf{q}n}^{RPA}=\frac{v_{\mathbf{q}}^{2D}\sinh \left[ qd\right] }{%
\sqrt{b_{\mathbf{q}n}^{2}-1}},  \label{inlayer repulsion}
\end{equation}%
where $b_{\mathbf{q}n}=\cosh \left( dq\right) -v_{\mathbf{q}}^{2D}\Pi _{%
\mathbf{q}n}\sinh \left( dq\right) $. This formula is reliable only away
from plasmons $b_{\mathbf{q}n}>1$. It turns out that to properly describe
superconductivity, one can simplify the calculation at low temperature by
considering the static limit $\Pi _{\mathbf{q}n}\simeq \Pi _{\mathbf{q}0}$.
Consequently the potential becomes static: $V_{\mathbf{q}}^{RPA}\equiv V_{%
\mathbf{q},n=0}^{RPA}$.

\section{Superconductivity}

Superconductivity in WSM is caused by a conventional phonon pairing. The
leading mode is an optical phonon mode assumed to be dispersionless with
energy $\Omega $. The effective electron-electron interaction due to the
electron - phonon attraction opposed by Coulomb repulsion (pseudo -
potential) creates pairing below $T_{c}$. Further we assume the singlet $s$%
-pairing channel and neglect the interlayer electrons pairing. \ It
important to note that unlike in conventional 3D metal superconductors where
a simplified pseudo - potential approach due to McMillan and other \cite%
{McMillan}, in 2D and layered WSM, one have to resort to a more microscopic
approach.

\subsection{Effective attraction due to phonon exchange opposed by the
effective Coulomb repulsion}

The free and the interaction parts of the effective electron action
("integrating phonons"+RPA Coulomb interaction) \cite{Rosenstein21} in the
quasi-momentum - Matzubara frequency representation, $S=S^{e}+S^{int}$,

\begin{eqnarray}
S^{e} &=&\frac{1}{T}\sum \limits_{\mathbf{k},l,n}\psi _{\mathbf{k}ln}^{\ast
sA}\left \{ \left( -i\omega _{n}+d_{\mathbf{k}}^{0}\right) \delta
^{AB}+\sigma _{i}^{AB}d_{\mathbf{k}}^{i}\right \} \psi _{\mathbf{k}ln}^{sB}%
\text{;}  \label{Interaction action} \\
S^{int} &=&\frac{1}{2T}\sum \nolimits_{\mathbf{q}nn^{\prime }mm^{\prime }}n_{%
\mathbf{q}ln}\left( \delta _{ll^{\prime }}V_{\mathbf{q,}m-m^{\prime
}}^{ph}+V_{\mathbf{q},l-l^{\prime }}^{RPA}\right) n_{-\mathbf{q},-l^{\prime
},-n^{\prime }}\text{.}  \notag
\end{eqnarray}%
Here $n_{\mathbf{q}ln}=\sum \nolimits_{\mathbf{p}}\psi _{\mathbf{p}ln}^{\ast
sI}\psi _{\mathbf{q-p,}l,n}^{sI}$ the Fourier transform of the electron
density. The effective electron - electron coupling due to phonons is:

\begin{equation}
\text{ \ }V_{\mathbf{q}m}^{ph}=-\frac{g^{2}\Omega }{\omega _{m}^{b2}+\Omega
^{2}}\text{,}  \label{electron phonon plus Coulomb}
\end{equation}%
where the bosonic frequencies are $\omega _{m}^{b}=2\pi mT$.

The pressure dependence on the frequency is approximated as: 
\begin{equation}
\Omega \left( P\right) =\Omega _{a}\left( 1+\zeta P\right) .
\label{Omega pressure}
\end{equation}%
For $HfTe_{5}$ we take $\Omega _{a}=15meV$ and $\zeta =0.005/GPa$.

\subsection{Nambu Green's functions and Gorkov equations}

Normal and anomalous (Matsubara) intra layer Nambu Green's functions are
defined by expectation value of the fields, $\left \langle \psi _{\mathbf{k}%
nl}^{Is}\psi _{\mathbf{k}nl}^{\ast s^{\prime }J}\right \rangle =\delta
^{ss^{\prime }}G_{\mathbf{k}n}^{IJ}$ and $\left \langle \psi _{\mathbf{k}%
nl}^{Is}\psi _{-\mathbf{k,-}n,l}^{Js^{\prime }}\right \rangle =\varepsilon
^{ss^{\prime }}F_{\mathbf{k}n}^{IJ}$, while the gap function is%
\begin{equation}
\Delta _{\mathbf{q}n}^{IJ}=\sum \nolimits_{\mathbf{p}m}V_{\mathbf{q-p,}%
n-m}F_{\mathbf{p}m}^{IJ},  \label{deltadef}
\end{equation}%
\begin{figure}[h]
\centering \includegraphics[width=12cm]{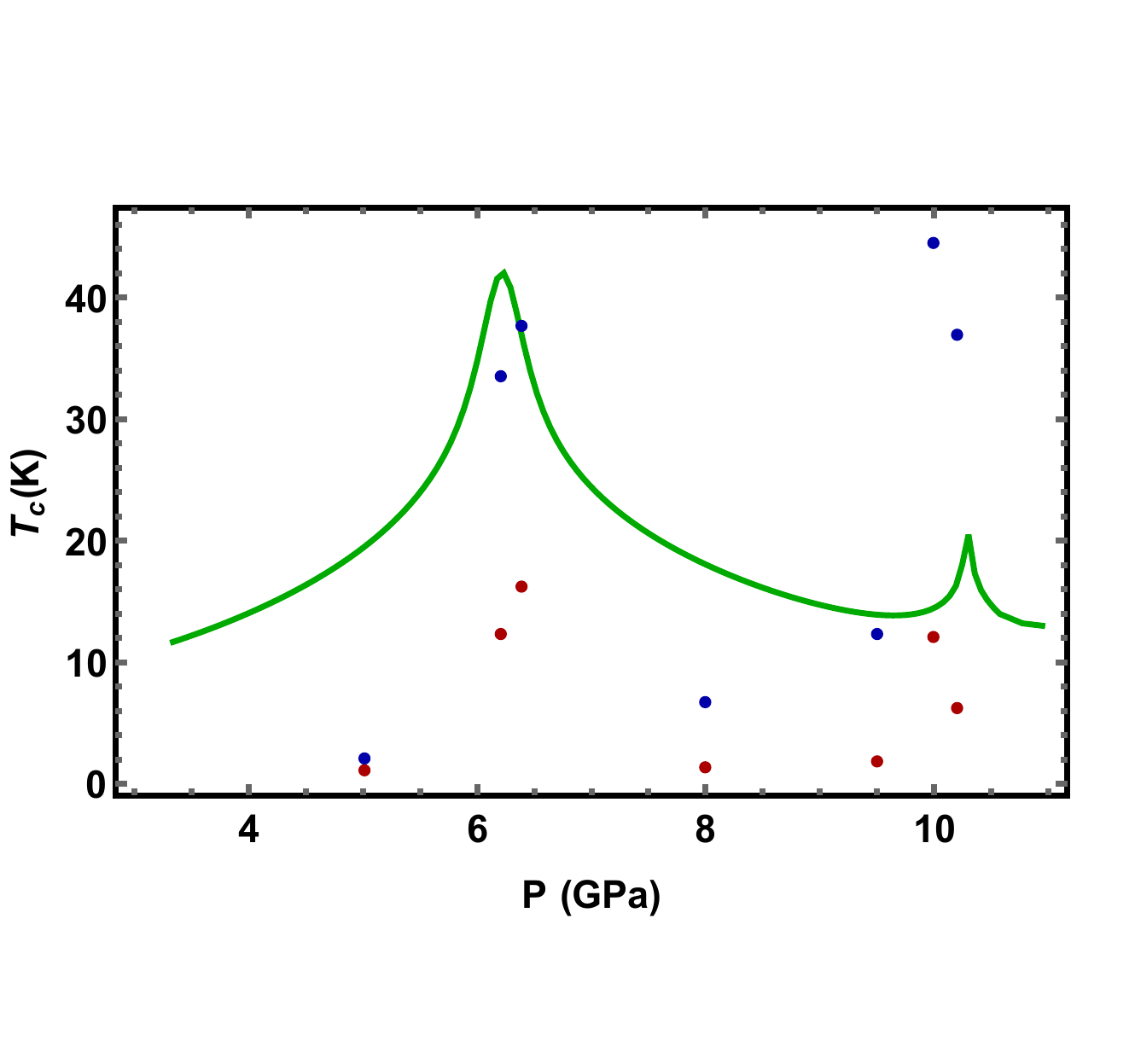}
\caption{The critical temperature $T_{c}$ as function of the hydrostatic
pressure $P$ with (red points) and without (blue points) the Coulomb
electron-electron interaction. The dependence has spikes near the points of
topological transformations of the electronic system. Position of spikes
coinsides with that of the density of states (the green curve).}
\end{figure}
where $V_{\mathbf{q}n}=V_{\mathbf{q}n}^{ph}+V_{\mathbf{q}n}^{RPA}$ is a
sublattice scalar. The gap equations in the sublattice matrix form are
derived from Gorkov equations\cite{Rosenstein21}:

\begin{equation}
\Delta _{\mathbf{q}n}=-\sum\nolimits_{\mathbf{p}m}V_{\mathbf{q-p,}n-m}g_{%
\mathbf{p}m}\left\{ I+\Delta _{\mathbf{p}m}g_{-\mathbf{p},-m}^{t}\Delta _{-%
\mathbf{p},-m}^{\ast }g_{\mathbf{p}m}\right\} ^{-1}\Delta _{\mathbf{p}m}g_{-%
\mathbf{p},-m}^{t}\text{.}  \label{Gap}
\end{equation}
This equation was solved numerically by iterations method. The momenta are
discretized as $q_{1.2}$ $=2\pi j_{1,2}/N_{s}$ $\ $(where $j_{1,2}$ $=$ $%
-N_{s}/2$ ...$\left( N_{s}/2-1\right) $) $N_{s}=256$ while the frequency
cutoff was $N_{T}=128$ the interatomic in-plane distance $a=3.5A$,
electron-phonon coupling $g\allowbreak \allowbreak =140meV$ and the
dielectric constant $\varepsilon =20.$

The critical temperature as a function on the pressure is presented in
Fig.3. The blue points represent the $T_{c}$ when the Coulomb repulsion is
neglected. It clearly shows the spikes of the $T_{c}$ near the points of the
both topological transformation of the electronic system caused by the
hydrostatic pressure. It amplifies the dependence of the density of states
(green line) in these points that can be understood from the approximate
exponential BCS dependence, $T_{c}=\Omega e^{-D\left( \mu \right) g^{2}}$. A
more realistic model includes the Coulomb repulsion, see red points in
Fig.3. The critical temperatures are much smaller demonstrating that in the
present case the repulsion plays the essential role. It turns out that it
not possible to approximate this behavior using a simplistic pseudo -
potential approach by McMillan\cite{McMillan} theory successfully applied to
3D good metals.

\section{Conclusion}

To summarize we have developed a theory of superconductivity in layered Weyl
semi-metals under the hydrostatic pressure that properly takes into account
the Coulomb repulsion. It is shown that in Weyl semi - metals the tilt of
the dispersion relation cones, $\kappa ,$ increases with pressure, so that
one is able to reach type II ($\kappa >1$starting from the more conventional
type I Weyl semi - metals , $\kappa <1$). It is found that upon increasing
pressure the I to II transition occurs in two continuous steps. In the first
step the cones of opposite chirality coalesce so that the chiral symmetry is
restored, while the second transition to the Fermi surface extending
throughout the Brillouin zone occurs at higher pressures. \ We show that the
critical temperature is a very robust tool to study these transformations of
the electronic system. The critical temperature shows spike in the points of
topological transformation repeating the density of the electron states. The
generalization goes beyond the simplistic pseudo - potential approach by
McMillan\cite{McMillan} theory. Superconductivity demonstrated significant
effect of the Coulomb repulsion on the critical temperature.

Acknowledgements. This work was supported by NSCof R.O.C.Grants
No.101-2112-M-009-014-MY3.

\bigskip

\bigskip


\bigskip

\begin{thebibliography}{99}
\bibitem{Katzenelson} \ \bigskip Katsnelson M.I. , \textit{The Physics of
Graphene}, Cambridge University Press, 2nd Edition (2012).

\bibitem{Weng} Weng H. , Dai X. and Fang Z. , Topological semimetals
predicted from first-principles calculations \ \textit{J. Phys. Cond. Matter.%
} \textbf{28}, 303001 (2016). Bansil A. , Lin H. , and Das T. , Colloquium:
Topological band theory \textit{Rev. Mod. Phys.} \textbf{88}, 021004
(2016);Weng H. , Fang C. , Fang Z. , Bernevig B. A. , and Dai X. , Weyl
semimetal phase in noncentrosymmetric transition-metal monophosphides, 
\textit{\ Phys. Rev. X\ 5, 011029 (2015)}; Lv B.Q. et al. Experimental
Discovery of Weyl Semimetal TaAs. \textit{Phys. Rev. X 5, 031013 (2015)};Xu
S.-Y. et al., Discovery of a Weyl fermion semimetal and topological Fermi
arcs, \textit{Science }\textbf{349}, 613 (2015).

\bibitem{MoTeearly} Huang L. et al., Spectroscopic evidence for a type II
Weyl semimetallic state in MoTe$_{2}$ \textit{Nature Materials }\textbf{15},
1155 (2016); Wang Y. et al, Gate-tunable negative longitudinal
magnetoresistance in the predicted type-II Weyl semimetal WTe$_{2}$ \textit{%
Nature Com. }\textbf{7}, 13142 (2016); Deng K. et al., Experimental
observation of topological Fermi arcs in type-II Weyl semimetal $MoTe_{2},$ 
\textit{Nature Physics} \textbf{12}, 1105 (2016).

\bibitem{ZrTe} Cao J. , et al., Landau level splitting in Cd$_{3}$As$_{2}$
under high magnetic fields, \textit{Nat. Comm.} \textbf{6}, 7779 (2015); Yu
W. Quantum Oscillations at Integer and Fractional Landau Level Indices in
Single-Crystalline ZrTe$_{5}$, \textit{Scientific Rep.} \textbf{6}, 35357
(2016).\ 

\bibitem{Armitage} Armitage N.P. , Mele E.J. and Vishwanath A., \textit{Rev.
Mod. Phys.} \textbf{90} 015001 (2018);Wang S. ,Lin B.-C. ,Wang A.-Q. ,Yu
D.-P. and Liao Z.-M. Quantum transport in Dirac and Weyl semimetals: a
review \textit{Adv. Phys. }X \textbf{2} 518-544 (2017);Yan B. and Felser C.,
Topological Materials: Weyl Semimetals, \textit{Annu. Rev. Condens. Matter
Phys.} \textbf{8} 337 (2017).

\bibitem{Goerbig} Katayama S. , Kobayashi A. , Suzumura Y. ,
Pressure-Induced Zero-Gap Semiconducting State in Organic Conductor
(BEDT-TTF)$_{2}$I$_{3}$ Salt, \textit{J. Phys. Soc. Japan} \textbf{75},
054705 (2006);Goerbig M. O. , Fuchs J. -N. , Montambaux G. , Pi\'{e}chon F., 
\textit{Tilted anisotropic Dirac cones in quinoid-type graphene and
(BEDT-TTF)}$_{2}$\textit{I}$_{3}$ Phys. Rev. B \textbf{78}, 045415 (2008);
Hirata M. et al, Observation of an anisotropic Dirac cone reshaping and
ferrimagnetic spin polarization in an organic conductor, \textit{Nature
Commun.} \textbf{7}, 12666 (2016).

\bibitem{Soluyanov} Soluyanov A. A. et al., Type-II Weyl semimetals, \textit{%
Nature} \textbf{527}, 495 (2015).

\bibitem{Yu} Yu Z.-M. ,Yao Y. , and Yang S. A. , Predicted Unusual
Magnetoresponse in Type-II Weyl Semimetals, \textit{Phys. Rev. Lett. }%
\textbf{117}, 077202 (2016).

\bibitem{Brien} O'Brien T. E. ,Diez M. , and Beenakker C. W. J. , Magnetic
Breakdown and Klein Tunneling in a Type-II Weyl Semimetal, \textit{Phys.Rev.
Lett.} \textbf{116}, 236401 (2016).

\bibitem{toptransition} Zhou Y. et al. Pressure-Induced New Topological Weyl
Semimetal Phase in TaAs, \textit{Phys. Rev. Lett.,} \textbf{117}, 146402
(2016).

\bibitem{CdAs} Liu Z.K. et al., A stable three-dimensional topological Dirac
semimetal Cd$_{3}$As$_{2}$, \textit{Nat.Mater.} \textbf{3} 677 (2014);
Borisenko S. et al. Experimental Realization of a Three-Dimensional Dirac
Semimetal, \textit{Phys.Rev.Lett.}\textbf{113} 027603 (2014); Neupane M. et
al, Observation of a three-dimensional topological Dirac semimetal phase in
high-mobility Cd$_{3}$As$_{2}$, Nat.Commun. \textbf{5} 3786 (2014).

\bibitem{NaBi} Liu Z.K. et al, Discovery of a Three-Dimensional Topological
Dirac Semimetal, $Na_{3}Bi$, \textit{Science} \textbf{343} 864 (2014).

\bibitem{Wang} Wang, C. et al. Monolayer atomic crystal molecular
superlattices, \textit{Nature}, \textbf{555}, 231 (2018); Lin, Z. et al.
Solution-processable 2D semiconductors for high-performance large-area
electronics, \textit{Nature} \textbf{562}, 254 (2018); Dresselhaus, M. \
Dresselhaus, G., Intercalation compounds of graphite, \textit{Adv. in Phys.}%
, \textbf{30} 139 \ (1981); Huang H.,Zhou S. ,and Duan W., Type-II Dirac
fermions in the $PtSe_{2}$ class of transition metal dichalcogenides \textit{%
Phys.Rev.B} \textbf{94},121117 (2016); Yan M. et al. Lorentz-violating
type-II Dirac fermions in transition metal dichalcogenide $PtTe_{2}$, 
\textit{Nature Comm.} \textbf{8}, 257(2017); Furue Y. Superconducting and
structural properties of the type-I superconductor $PdTe_{2}$ under high
pressure, \textit{Phys. Rev. B} \textbf{104}, 144510 (2021);

\bibitem{Britnel} Duong D. L. ,Yun S. J. , and Lee Y. H. , van der Waals
Layered Materials: Opportunities and Challenges, \textit{ACS Nano}, \textbf{%
11} 11803 (2017).

\bibitem{TaX2} Adam M. L. and Bala A. A. , Superconductivity in quasi-2D $%
InTaX_{2}$ (X = S, Se) type-II Weyl semimetals, J. Phys.:Condens.Matter 
\textbf{33 }225502 (2021).

\bibitem{pressureTc} Deng W. et al Pressure-Quenched Superconductivity in
Weyl Semimetal NbP Induced by Electronic Phase Transitions under Pressure, 
\textit{J. Phys. Chem. Lett.} \textbf{13}, 5514\ (2022); van Delft M. R., et
al, Two- and Three-Dimensional Superconducting Phases in the Weyl Semimetal
TaP at Ambient Pressure, \textit{Crystals} \textbf{10}, 288 (2020).

\bibitem{PdTe2} Xiao R.C.,et al. Manipulation of type-I and type-II Dirac
points in $PdTe_{2}$ superconductor by external pressure, Phys.Rev. B 
\textbf{96}, 075101 (2017); Leng H.,et al. Superconductivity under pressure
in the Dirac semimetal $PdTe_{2}$ \textit{J. Phys.: Condens.Matter} \textbf{%
32},025603 (2020);Yang H. et al., Anomalous charge transport of
superconducting $Cu_{x}PdTe_{2}$ under high pressure, \textit{Phys.Rev. B} 
\textbf{103}, 235105 (2021).

\bibitem{ZrTeZhou} Zhou Y. et al., Pressure-induced superconductivity in a
three-dimensional topological material $ZrTe_{5}$ \textit{\ PNAS}, \textbf{15%
} \ 2904-2909 \ (2016).

\bibitem{Liu17} Liu Y. et al. Superconductivity in HfTe$_{5}$ across weak to
strong topological insulator transition induced via pressures, Scientific
Reports, \textbf{7} 44367 (2017).

\bibitem{Kusmartsev} Hills R.D.Y., Kusmartseva A. and Kusmartsev F.V.,
Current-voltage characteristics of Weyl semimetal semiconducting devices,
Veselago lenses, and hyperbolic Dirac phase, \textit{Phys. Rev. B} \textbf{95%
} 214103 (2017).

\bibitem{Yeh} Sun F. and Ye J. , Type-I and type-II Weyl fermions,
topological depletion, and universal subleading scaling across topological
phase transitions, \textit{Phys. Rev. B} \textbf{96}, 035113 (2017).

\bibitem{Rosenstein17} Li D. ,Rosenstein B. ,Shapiro B. Ya. , and Shapiro
I., Effect of the type-I to type-II Weyl semimetal topological transition on
superconductivity, \textit{Phys. Rev. B \ }\textbf{95}, 094513 (2017).

\bibitem{Zyuzin} Alidoust M. ,Halterman K. , and Zyuzin A. A. ,
Superconductivity in type-II Weyl semimetals, \textit{Phys. Rev. B} \ 
\textbf{95}, 155124 (2017).

\bibitem{Rosenstein18} Li D. ,Rosenstein B. ,Shapiro B. Ya. , and Shapiro I.
, Magnetic properties of type-I and type-II Weyl semimetals in the
superconducting state, \textit{Phys. Rev. B} \textbf{97}, 144510 (2018).

\bibitem{GL18} Rosenstein B. , Shapiro B Ya , Li D. , Shapiro I., Upper
critical magnetic field in superconducting Dirac semimetal, \textit{%
Europhys. Lett.} \textbf{124} 27004 \ (2018).

\bibitem{Li15} Lee C. - H. et al. \textit{Sci. Rep.} \textbf{5}, 10013
(2015).

\bibitem{McMillan} Bilbro G. and McMillan L. , Theoretical model of
superconductivity and the martensitic transformation in A15 compounds, 
\textit{Phys. Rev. B} \textbf{14 }1887 (1976).

\bibitem{Elliasson} Hawrylak P. ,Eliasson G. , and Quinn J. J., Many-body
effects in a layered electron gas, \textit{Phys. Rev. B} \textbf{37} 10187
(1988).

\bibitem{grapheneoxide} Didenkin A.T. and A.Y. Vul, Graphene Oxide and
Derivatives: The Place in Graphene Family, \textit{Frontiers in Physics, }%
\textbf{6}, 1 (2019)

\bibitem{Goerbig2} Monteverde M. et al. Coexistence of Dirac and massive
carriers in (BEDT-TTF)$_{2}$I$_{3}$ under hydrostatic pressure, \textit{%
Phys. Rev B} \textbf{87}, 245110 (2013).

\bibitem{Sun} Sun Y. , Wu S.-C. , Ali M. N., Felser C. , and Yan B. ,
Prediction of Weyl semimetal in orthorhombic $MoTe_{2}$ , \textit{\ Phys.
Rev.} B \textbf{92}, 161107(R) (2015); Ruan J. et al., Symmetry-protected
ideal Weyl semimetal in HgTe-class materials, \textit{Nature Com.} \textbf{7}
11136 (2016).

\bibitem{MoTe2} Deng K. et al., Experimental observation of topological
Fermi arcs in type-II Weyl semimetal $MoTe_{2}$, \textit{Nat. Phys.} \textbf{%
12 }1105 (2016).

\bibitem{Volovik} Volovik G.E. \ Exotic Lifshitz transitions in topological
materials, \textit{Phys.-Usp.} \textbf{61,} 89 \ (2018).

\bibitem{dielectric} Beal A R and Hughes H P, Kramers-Kronig analysis of the
reflectivity spectra of $2H-MoS_{2}$, $2H-MoSe_{2}$ and $2H-MoTe_{2}$, 
\textit{J. Phys. C: Solid State Phys.} \textbf{12,} 881 (1979).

\bibitem{Rosenstein21} Rosenstein B. and Shapiro B. Ya., Apical oxygen
vibrations dominant role in d-wave cuprate superconductivity and its
interplay with spin fluctuations, \textit{J. Phys. Commun.} \textbf{5}
055013 (2021).

\bibitem{Zhang22} Zhang H. 2D Mater.\textbf{\ }Enhanced superconductivity
with interlayer spacing dependent Tc in intercalated Weyl semimetal $%
MoTe_{2} $\textbf{\ 9} 045027 (2022)
\end{thebibliography}
\end{document}